\documentclass [amssymb,amsmath,twocolumn, showpacs]{revtex4-1} % twocolumn, 
\usepackage{graphicx,epsfig,amsfonts,amssymb}
\usepackage{bm}% bold math
\usepackage{times}
\usepackage{lipsum}

\newcommand{\beq}{\begin{equation}}
\newcommand{\eeq}{\end{equation}}
\newcommand{\bes}{\begin{subequations}}
\newcommand{\ees}{\end{subequations}}
\newcommand{\bea}{\begin{eqnarray}}
\newcommand{\eea}{\end{eqnarray}}
\newcommand{\ba}{\begin{array}}
\newcommand{\ea}{\end{array}}
\newcommand{\beqn}{\begin{eqnarray*}}
\newcommand{\eeqn}{\end{eqnarray*}}

\newcommand{\ra}{\rangle}

\newcommand{\rar}{\rightarrow}
\newcommand{\upa}{\uparrow}
\newcommand{\dna}{\downarrow}

\begin{document}

\title{Solvable 4-state Landau-Zener model of two interacting qubits with path interference}

\author {N.~A. {Sinitsyn}$^{a}$ }
\address{$^a$ Theoretical Division, Los Alamos National Laboratory, Los Alamos, NM 87545,  USA}

\begin{abstract}
We identify a nontrivial 4-state Landau-Zener model for which transition probabilities between any pair of diabatic states can be determined analytically and exactly. The model describes an experimentally accessible system of two interacting qubits, such as a localized state in a Dirac material with both valley and spin degrees of freedom or a singly charged quantum dot (QD) molecule with  spin orbit coupling. Application of the linearly time-dependent magnetic field induces a sequence of quantum level crossings with possibility  of interference of different  trajectories in a semiclassical picture. We argue that this system satisfies the criteria of integrability in the multistate Landau-Zener theory, which allows us to derive explicit exact analytical expressions for the  transition probability matrix. We also argue that this model is likely a special case of a larger class of solvable systems, and present a 6-state generalization as an example. 
\end{abstract}
\date{\today}

\maketitle

%\section{Introduction}
 
%Explicitly time-dependent Schr\"odinger equation is
%of importance for numerous applications. 
%Quantum mechanical nonadiabatic transitions have been studied in molecular collisions for long time \cite{rozen, nikitin, osherov}.
%Relatively recently, 

%Multichannel nonadiabatic processes are frequently found in control of quantum many-body states in electronics, magnetic systems, and Bose condensates \cite{app-el,app-bose,app-spin,app-exp}.  
 %Condensed matter physicists often achieve an insight about their systems by finding a proper model from a large variety of exactly solvable systems with well defined thermodynamic limit. Unfortunately, most of such known models correspond to the  stationary  Schr\"odinger equation. Exact solutions of explicitly time-dependent systems with a large number of states are  rare. 
 %They are also often trivially reducible to independent oscillators or non-interacting spins. 
 %It becomes important for the future progress of this field to explore new integrable models of quantum mechanical evolution with time-dependent parameters and with a large, possibly macroscopic, number of interacting states. 
 %%%%%%%%%%%%%%%%%%%%%%%%%%%%%%%%%%%%%%%%%%%%%%%%%%%%%%%%%%%%%%%%%%%%%%%%%%%%%%%%%%%%%%%%%%%
 
% \cite{book,maj,landau, LZ, stuck}

\section{Introduction}

Measurements of Landau-Zener transition probabilities have become a powerful tool for characterization of coupling constants and quantum state preparation in quantum dots (QDs) \cite{dot-lz-exp1,multiparticle} and QD-molecules \cite{dot-lz-exp2,dot-lz-theory1}.
%molecular nanomagnets \cite{app-spin}, and confined ultracold atomic systems \cite{atomic}.  Landau-Zener transitions are also frequently discussed in the context of quantum control of nanoscale systems \cite{qcontrol,app-bose} and coherence \cite{coher}. A growing range of applications has lead to emergence of a new measurement approach called the Landau-Zener interferometry \cite{LZ-interferometry}.
The central theoretical result, which is frequently used in these studies, is the Landau-Zener formula \cite{book} that provides transition probabilities for a 2-level system with a time-dependent linear diabatic level crossing. However, most of the modern applications  consider  interactions of more than two states. In such cases, even linear time-dependence of parameters does not usually implies the possibility to derive exact analytical predictions. While numerical simulations of such systems are accessible for a small number of interacting states, exact nonperturbative analytical results are  desirable to develop the intuition and explore the possibility of optimization of quantum dynamics. 

The multistate version of the two-state Landau-Zener model is one of the most fundamental systems in nonstationary quantum mechanics \cite{be}.
It describes the evolution of $N$  states according to the Sch\"odinger equation with parameters that change linearly with time:
\begin{equation}
i\frac{d\psi}{d t} = \left(\hat{A} +\hat{B}t  \right)\psi.
\label{mlz}
\end{equation} 
Here, $\psi$ is the state vector in a space of $N$ states; $\hat{A}$ and $\hat{B}$ are constant Hermitian $N\times N$ matrices.  One can always choose the, so-called, {\it diabatic basis} in which the matrix $\hat{B}$ is diagonal,
and if any pair of its elements are degenerate then the corresponding off-diagonal element of the matrix $\hat{A}$ can be set to zero, that is
\beq
B_{ij}= \delta_{ij}\beta_i, \quad  A_{nm}=0\,\,\, {\rm if} \,\, \beta_{n}=\beta_{m},\,\,n\ne m \in (1,\ldots N).
\label{diab1}
\eeq
Constant parameters $\beta_i$  are called the {\it slopes of diabatic levels};  nonzero off-diagonal elements of the matrix $\hat{A}$ in the diabatic basis are called the {\it coupling constants},  and the diagonal elements of the Hamiltonian, 
$$
H_{ii}=\beta_i t +\epsilon_i,
$$
 where $\epsilon_i \equiv A_{ii}$, are called the {\it diabatic energies}. The goal of the multistate Landau-Zener theory is to find the scattering $N\times N$ matrix $\hat{S}$, whose element $S_{nn'}$ is the amplitude of the diabatic state $n$ at $t  \rightarrow +\infty$, given that at $t \rightarrow -\infty$ the system was in the $n'$-th diabatic state. In most cases, only the related matrix $\hat{P}$, $P_{nn'}=|S_{nn'}|^2$,  called the {\it matrix of transition probabilities}, is of interest. 
Generally, for $N>2$, the analytical solution of the model (\ref{mlz}) is unknown. Nevertheless, a number of exactly solvable cases with specific forms of matrices $\hat{A}$ and $\hat{B}$  have been derived \cite{multiparticle, do, bow-tie, six-LZ,armen}. 

In this article, we present a new solvable system of the type (\ref{mlz}) which can be realized in experiments on Landau-Zener transitions in quantum dots and quantum dot molecules \cite{dot-lz-exp1,dot-lz-exp2}. 
It describes an interaction of a pair of two-level systems (qubits) with a linearly time-dependent magnetic field. 

\section{Model}
Let $\hat{\sigma}_{\alpha}$ and $\hat{s}_{\alpha}$, $\alpha=x,y,z$, be the Pauli matrices 
acting in the space of, respectively, the first and the second qubit. %We will also use raising/lowering operators $\hat{s}^{\pm}$ in the  space of the second qubit.
The most general Hamiltonian that we consider reads:

\beq
  \hat{H} =E \hat{\sigma}_z  + g \hat{\sigma}_x +\gamma \hat{\sigma}_y \hat{s}_y+\beta_1 t \hat{s}_z +\beta_2 t \hat{\sigma}_z \hat{s}_z. 
 \label{h-pauli1}
 \eeq
 Physically, this Hamiltonian describes a QD-molecule made of two small QDs charged with a single electron (or a hole) that can tunnel between them. 
Operators $\hat{s}_{\alpha}$ act in the space of the true spin of the electron and operators $\hat{\sigma}_{\alpha}$ act in the space of two spatially separated localized states in the QD-molecule.
Thus, the first term in (\ref{h-pauli1}) describes the energy difference, $2E$, between the QDs in the limit of zero coupling between them. The second term describes spin-conserving charge tunneling with amplitude $g$. 
The third term describes the tunneling between QDs that  flips the spin. This  term is expected in the presence of the spin orbit coupling. For example, heavy-hole states of
different InGaAs quantum dots generally have different degrees of mixing to the light-hole states. This difference originates from different shapes of self-assembled QDs. Therefore, the tunneling hole  experiences different spin-orbit field in different QDs.  The form of the third term in (\ref{h-pauli1}) is constrained  by the time-reversal symmetry at $t=0$. 
 
 The last two terms in (\ref{h-pauli1}) describe interaction of the QD-molecule with an external linearly time-dependent out-of-plane magnetic field: the term with $\beta_1$ describes the average Zeeman coupling, while the last term, with $\beta_2$ in (\ref{h-pauli1}), originates from the difference of the Lande g-factors of different QDs.

%%%%%%%%%%%%%%%%%%%%%%%%%%%%%%%%%%%%%%%%%%%%%%%%%%%%%%%%%%%%%%%%%%%%%%%%%%%%%%%%%%%%%%%%%%%
\begin{figure}%[!htb]
\scalebox{0.43}[0.43]{\includegraphics{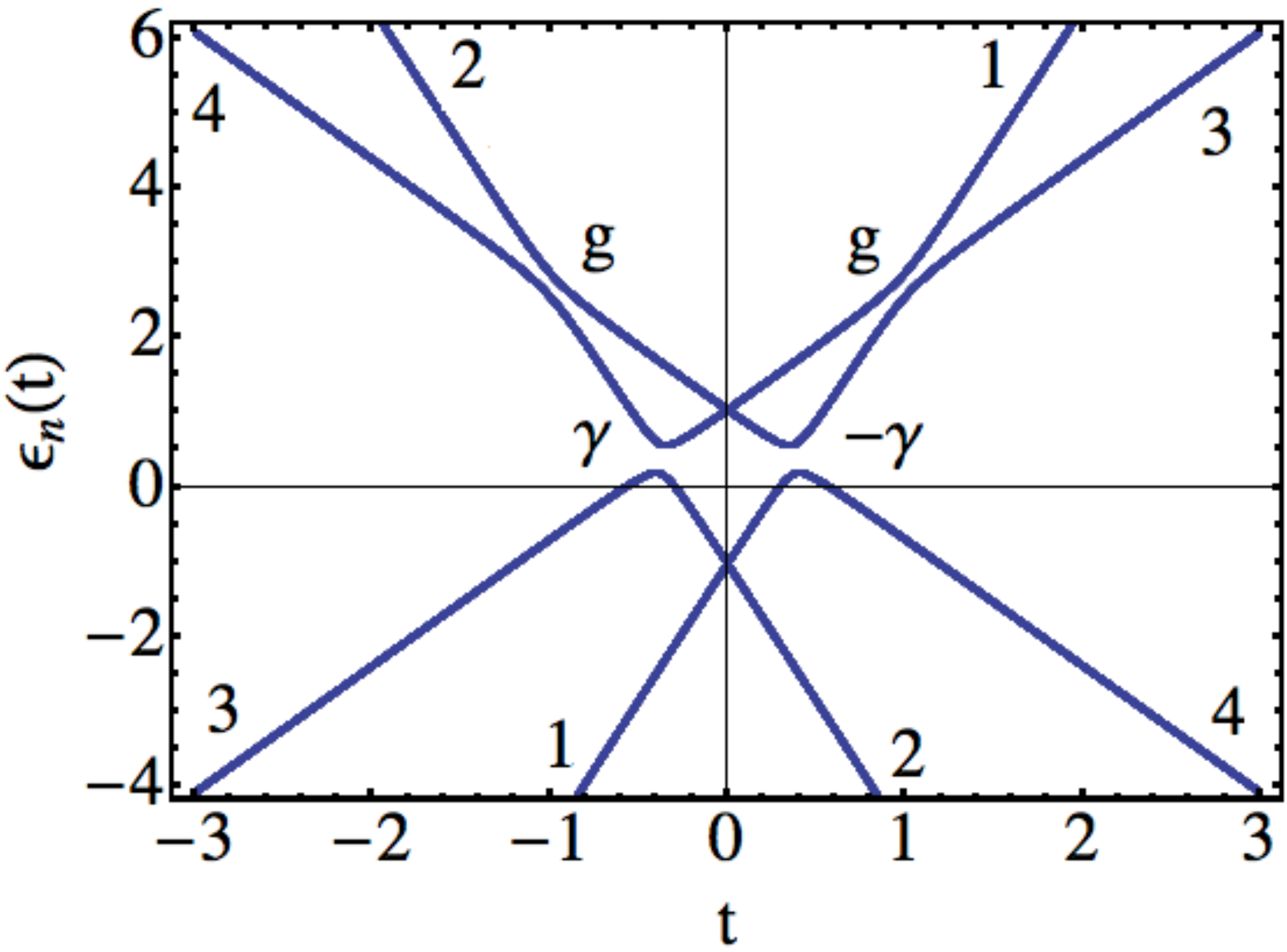}}
\hspace{-2mm}\vspace{-4mm}   
\caption{ (Color online) Adiabatic energy levels (blue curves) and their pairwise couplings (parameters $g$ and $\pm \gamma$) of the Hamiltonian (\ref{ham1}) shown at corresponding avoided level crossings. Ket-vectors mark  states that correspond to adiabatic energies at $t\rar -\infty$. Two exact crossing points at $t=0$ are guaranteed by the time-reversal symmetry. Parameters are: $\beta_1=2.7$, $\beta_2=1.0$, $E=-1$, $g=0.16$, $\gamma=0.2$.}
\label{levels}
\end{figure}
%%%%%%%%%%%%%%%%%%%%%%%%%%%%%%%%%%%%%%%%%%%%%%%%%%%%%%%%%%%%%%%%%%%%%%%%%%%%%%%%%%%%%%%%%%%
 
In the basis, $(1,2,3,4)\equiv ( |\upa \upa \ra, |\upa \dna \ra, |\dna \upa \ra,  |\dna \dna \ra)$, of eigenstates of operator $\hat{\sigma}_z \hat{s}_z$,  the Hamiltonian (\ref{h-pauli1}) is a 4$\times$4 matrix:
\begin{equation}
\hat{H}=\left( 
\begin{array}{cccc}
b_1 t +E & 0 & g & -\gamma  \\
0 & -b_1 t +E & \gamma & g  \\
g & \gamma & b_2 t-E & 0 \\
-\gamma&g& 0& -b_2 t -E
\end{array}
\right),
\label{ham1}
\end{equation} 
where 
\begin{equation}
 b_1=\beta_1+\beta_2, \quad  \, b_2=\beta_1-\beta_2. 
\end{equation}
The Hamiltonian (\ref{ham1}) has already encountered in the literature on quantum dot molecules \cite{roy}. It is the most general, irreducible by a change of the basis, single electron  Hamiltonian for spin tunneling between two discrete states that is consistent with the time-reversal symmetry at zero magnetic field. We note here that the assumption of an out-of-plane direction of the magnetic field is not crucial. For the case of a tilted linearly time-dependent magnetic field, one can switch to the diabatic basis in which the Hamiltonian has a canonical form (\ref{ham1}) with redefined coupling constants. 

It is convenient to visualize a multistate Landau-Zener model with a plot of instantaneous eigenenergies of the Hamiltonian as functions of $t$, as shown in Fig.~\ref{levels} for the case of the Hamiltonian (\ref{ham1}). Note that due to the time-reversal symmetry at $t=0$ this plot contains two points of exact adiabatic level crossing. 

More exotic but physically possible realization of the Hamiltonian (\ref{ham1}) can also be  a single electron state localized near a short range impurity in a massive Dirac semiconductor, such as MoS$_2$. 
A short range potential can couple nearly degenerate states of different valleys. A localized state would be characterized then by both spin and valley indexes. Let $\hat{\tau}_{\alpha}$ be the Pauli operators acting in the valley space. Then 
the spin-valley Hamiltonian that is consistent with the time-reversal symmetry would be
\beq
  \hat{H}_{\rm sv} =E \hat{\tau}_z \hat{s}_{z}  + g \hat{\tau}_x - \gamma \hat{\tau}_z \hat{s}_x, 
 \label{h-pauli2}
 \eeq
 where the first term is due to the Kane-Mele type of  the spin-orbit coupling, the second term is due to the valley mixing by the impurity potential, and the third term is due to the Rashba-type of the spin-orbit coupling. 
Application of an out-of-plane magnetic field would introduce similar time-dependent terms $\beta_1t \tau_z+ \beta_2 t \hat{s}_z $, where  the term with $\beta_1$ describes the valley splitting by a magnetic field, the term with $\beta_2$ is the standard Zeeman coupling. 
In the basis, $ |\upa \upa \ra, |\dna \dna \ra, |\dna \upa \ra, |\upa \dna \ra$, of eigenstates of the operator $\hat{\tau}_z \hat{s}_z$, the Hamiltonian has the same matrix form as in Eq.~(\ref{ham1}).

As the coupling parameters for real QDs are expected to be in GHz range, it would be practically impossible to achieve the nonadiabatic regime of the model (\ref{ham1}) with  true time-dependent magnetic fields. However, optically induced {\it effective} magnetic fields with needed characteristics to control the quantum dot spins have been already demonstrated in self-assembled QD-molecules \cite{qd-molecule}. The idea of such an optical control is to couple the states of the quantum dot to a high-energy exiton state by a circularly polarized  off-resonant optical beam. If the frequency of the beam is well detuned, one can avoid dissipative transitions while inducing effective time-dependent Zeeman couplings via the inverse Faraday effect (also known as the optical Stark effect). Such optical pulses can be prepared with picosecond precision and considerable amplitude of the effective field that they create \cite{inverseFA-fast}. For example, valley selective energy splitting, as large as 18meV,  in WS$_2$ Dirac semiconductor was achieved in \cite{inverseFA-dirac}, which is similar to the expected value of the Kane-Mele spin orbit coupling of electronic bands of this material. 

We recall that the inverse Faraday effect appears as the 2nd order perturbation in the electric field, so the induced effective field is generally proportional to the intensity of the beam. 
For example, one can initialize the system in the ground state in the presence of a strong but constant magnetic field and apply an optical pulse with a linearly growing intensity so that the effective field sweeps from large negative to large positive values. After this, the pulse can be instantly terminated in presence of the original constant external field. Since, this would be equivalent to the sudden change of the field direction, such an optical pulse switch-off would not induce additional transitions among diabatic states. 
%Then the probabilities of final states can be measured by standard means.

% which should be sufficient to test our predictions.   

%This completes our discussion of a physical implementation of the model (\ref{ham1}). In what follows, we turn to the exact solution for the corresponding scattering problem. 

\section{Solution of the model}

\begin{figure}%[!htb]
\scalebox{0.10}[0.10]{\includegraphics{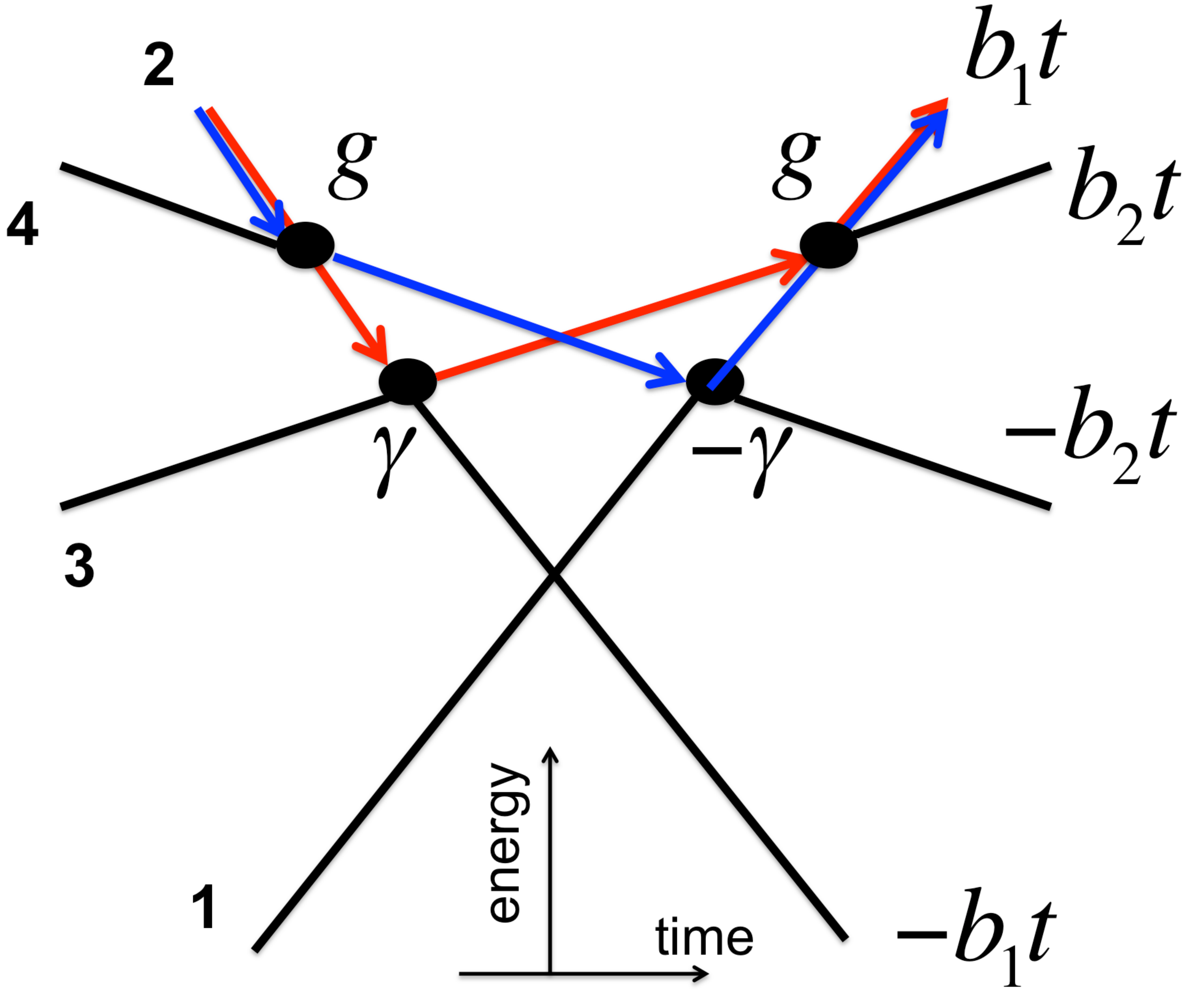}}
\hspace{-2mm}\vspace{-4mm}   
\caption{ (Color online) Diabatic energy levels (diagonal elements of the Hamiltonian (\ref{ham1})) and two semiclassical trajectories connecting the state $2\equiv |\upa \dna \ra$ at $t \rightarrow -\infty$ to the state $1 \equiv |\upa \upa \ra$ at $t \rightarrow +\infty$ (red and blue arrows). }
\label{diab}
\end{figure}

At large negative or positive times, eigenenergies of the Hamiltonian are well separated so that transitions between them are suppressed according to the adiabatic theorem. Adiabaticity is violated when 
pairs of energy levels appear close to each other. 
There are four avoided crossing points in Fig.~\ref{levels}. If such regions were sufficiently far from each other, one would be able to justify the, so-called, {\it independent crossing approximation}, according to which, in order to determine the scattering amplitudes, one should simply follow the diabatic levels, as illustrated in Fig.~\ref{diab}, and apply the two-state Landau-Zener formula to  pair-wise avoided level crossings in the chronological order of their appearance and trivially include the phase gain along semiclassical trajectories. 

It has been noticed previously \cite{multiparticle,do,be,bow-tie,six-LZ} that, surprisingly, all known exactly solvable models of the type (\ref{mlz}) with a finite number of interacting states have exact solutions for the scattering matrix that coincide with the  prediction of the independent crossing approximation. Moreover, all such solvable models have two properties \cite{six-LZ}: 

 (i) the absence of the dynamic phase effect on transition probabilities in the semiclassical framework, and 
 
 (ii) the exact crossing of adiabatic energies at intersections of diabatic states (i.e. diagonal elements of the Hamiltonian) without direct couplings. 

Here by the dynamic phase we mean the trivial phase 
\beq
\phi_{\rm dyn} = -\int_{-\infty}^{\infty} [ \beta_{k(t)} t +\epsilon_{k(t)}] \, dt,  
\label{dynph}
\eeq
where $k(t)$ is the index of the level along a semiclassical trajectory. Its time-dependence follows from the possibility of switching   diabatic level indexes at level crossing points.

The main observation of this article, is that the model (\ref{ham1}) satisfies criteria (i-ii). Consider Fig.~\ref{diab} that shows time-dependence of the diabatic energy levels in this model. Condition (ii) is satisfied trivially: diabatic levels 1 and 2 are not directly coupled and their crossing corresponds to the adiabatic level crossing in Fig.~\ref{levels}. The same is true for the case of the crossing of levels 3 and 4.

%%%%%%%%%%%%%%%%%%%%%%%%%%%%%%%%%%%%%%%%%%%%%%%%%%%%%%%%%%%%%%%%%%%%%%%%%%%%%%%%%%%%%%%%%%%
%%%%%%%%%%%%%%%%%%%%%%%%%%%%%%%%%%%%%%%%%%%%%%%%%%%%%%%%%%%%%%%%%%%%%%%%%%%%%%%%%%%%%%%%%%%

In order to verify the condition (i), we will  derive the prediction for the transition probabilities in the independent crossing approximation explicitly. Only trajectories that turn forward in time  in Fig.~\ref{diab} are allowed in the semiclassical limit.
According to \cite{bow-tie,six-LZ}, the transition amplitudes at a crossing of a pair of states after a passage through a crossing point are described by the Landau-Zener formula with a trivial assumption about the phase change. Namely, if a diabatic level $i$ crosses another level $j$ with a corresponding nonzero coupling $g_{ij}$ then one should assume that with an amplitude  
\beq
s_{i\rar i}=\sqrt{p_{i\rar j}}, \quad p_{i\rar j} \equiv e^{-2\pi |g_{ij}|^2|/|b_i -b_j|}
\label{lz1}
\eeq
 the system will remain in the same diabatic state after the level crossing. Here $p_{i\rar j}$ is the standard Landau-Zener transition probability for a two level crossing. 
 The amplitude  to turn to another level should be assumed
 \beq
s_{i\rar j}=\pm i \sqrt{1-p_{i\rar j}},
\label{lz2}
\eeq
where  $(\pm)$ is the sign of the coupling constant at the level intersection.  
 Here we note that the full semiclassical approximation predicts an additional complex phase prefactor in (\ref{lz2}) that depends on the coupling constants \cite{bow-tie}. However, in all found integrable models, this prefactor either cancels or factorizes from the final transition amplitude, and does not lead to the change of the transition probability matrix, as it is discussed in detail in \cite{bow-tie}. 
Therefore, we will assume the form of the phase prefactor as in Eq.~(\ref{lz2}) and test the final result a posteriory.  

Let us consider a transition from  level-1 at $t\rar -\infty$ to level-3 at $t\rar +\infty$.  Starting from state-1, we move along the diabatic level-1 in the positive time direction. 
First, according to Fig.~\ref{diab}, we encounter the crossing point with level-2, which is an exact crossing point that, according to (\ref{lz1}) does not influence dynamics in the independent crossing approximation. The next encountered crossing will be with  level-4 and coupling $-\gamma$. In order to end up on level-3 we have to assume that we pass this crossing by staying on level-1, which happens with the amplitude 
$$
s_{1 \rar 1} =e^{-\frac{\pi \gamma^2}{|b_1 -(-b_2)|}}=e^{-\frac{\pi \gamma^2}{2\beta_1}},
$$
where we assumed that $\beta_1,\beta_2>0$. The next encountered crossing point would be with level-3, at which we should turn. The corresponding amplitude is
$$
s_{1 \rar 3} =i\sqrt{1-e^{-\frac{\pi g^2}{\beta_2}}}.
$$
There are no more crossing points and no alternative trajectories connecting the same initial and final states. Therefore, we can write the semiclassical amplitude for the transition probability:
\beq
S_{31} = ie^{i\phi_{\rm dyn}^{31}} e^{-\frac{\pi \gamma^2}{2\beta_1}} \sqrt{1-e^{-\frac{\pi g^2}{\beta_2}}},
\label{s13}
\eeq
where $\phi_{\rm dyn}^{31}$ is the dynamic phase along this trajectory. The explicit expression for this phase is not needed because it cancels after we take the absolute value squared of the amplitude in order to obtain the transition probability from state-1 to state-3:
\beq
P_{31}=|S_{31}|^2 = e^{-\frac{\pi \gamma^2}{\beta_1}}\left( 1-e^{-\frac{\pi g^2}{\beta_2}} \right).
\label{p13}
\eeq
Thus, we found that the dynamic phase is canceled in the expression for the transition probability (\ref{p13}) obtained in the independent crossing approximation. Hence condition (i) is satisfied for this particular transition. 
The reason for cancellation of $\phi_{\rm dyn}^{31}$ is because there is only a single semiclassical trajectory describing a given transition, so there are no interference effects through which the dynamic phase can influence the transition probability.  By examining all possible transitions that start at states 1 and 3,  we find that the same property is satisfied for all of them. 

The only nontrivial situation is when the system starts either at  levels 2 or 4 and ends up at levels 1 or 3. An example for the transition from level-2 to level-1 is shown in Fig.~\ref{diab}, where we mark the allowed trajectories by red and blue arrows. One can find, however, that dynamic phases along the red and blue trajectories are identical. So, even though there is interference of two amplitudes, this particular phase factorizes and cancels in the expression for the transition probability from level-2 to level-1. Hence condition (i) is satisfied despite the path interference. To show this, we recall that the dynamic phase (\ref{dynph}) is the area between the trajectory and the time axis \cite{bow-tie}. Due to the symmetry of the diabatic level structure in Fig.~(\ref{diab}), under reflection $t\rar -t$ the blue and red trajectories transfer into each other under this transformation, which means that they sweep the same area below them. 
Hence, the dynamic phase  does not influence the final transition probability.

According to the rules (\ref{lz1})-(\ref{lz2}), we find amplitudes of the red and blue trajectories in Fig.~\ref{diab} up to the same dynamic phase factor:
\begin{eqnarray}
\label{red}
S_{12}^{\rm red} &=&  i^2 e^{-\frac{\pi g^2}{2\beta_2}}  \sqrt{(1-e^{-\frac{\pi \gamma^2}{\beta_1}} )(1-e^{-\frac{\pi g^2}{\beta_2}})} , \\
S_{12}^{\rm blue} &=&  -i^2 \sqrt{(1-e^{-\frac{\pi g^2}{\beta_2}})(1-e^{-\frac{\pi \gamma^2}{\beta_1}} )} e^{-\frac{\pi g^2}{2\beta_2}}.
\label{blue}
\end{eqnarray}

Comparing (\ref{red}) and (\ref{blue}), we find that the amplitudes of  red and blue trajectories are different only by a sign, which can be traced to the sign difference near the coupling $\gamma$ at two avoided crossings in Fig.~\ref{levels}. Therefore, these trajectories interfere destructively, and the total transition amplitude and the corresponding transition probability are identically zero:
\beq
P_{12}=0.
\label{p12}
\eeq

One can then find, e.g., that similar two trajectories that connect initial level-2 and final level-3 interfere constructively. We are now in a position to summarize predictions of the independent crossing approximation in the form of the matrix of transition probabilities, with elements $P_{nm}\equiv P_{m\rar n}$:
\beq
\hat{P} = \left( 
\begin{array}{cccc}
p_1p_2 &0 &p_2q_1 & q_2\\
0 & p_1p_2& q_2 & p_2q_1 \\
p_2q_1 & q_2 & p_1 p_2 & 0\\
q_2 & p_2 q_1 &0 & p_1 p_2
\end{array}
\right), 
\label{prob1}
\eeq
where
\beq
p_1\equiv e^{-\frac{\pi g^2}{\beta_2}}, \quad p_2 \equiv  e^{-\frac{\pi \gamma^2}{\beta_1}} \quad q_n\equiv 1-p_n.
\label{pq}
\eeq

\section{Numerical check}

%%%%%%%%%%%%%%%%%%%%%%%%%%%%%%%%%%%%%%%%%%%%%%%%%%%%%%%%%%%%%%%%%%%%%%%%%%%%%%%%%%%%%%%%%%%
\begin{figure}%[!htb]
\scalebox{0.24}[0.24]{\includegraphics{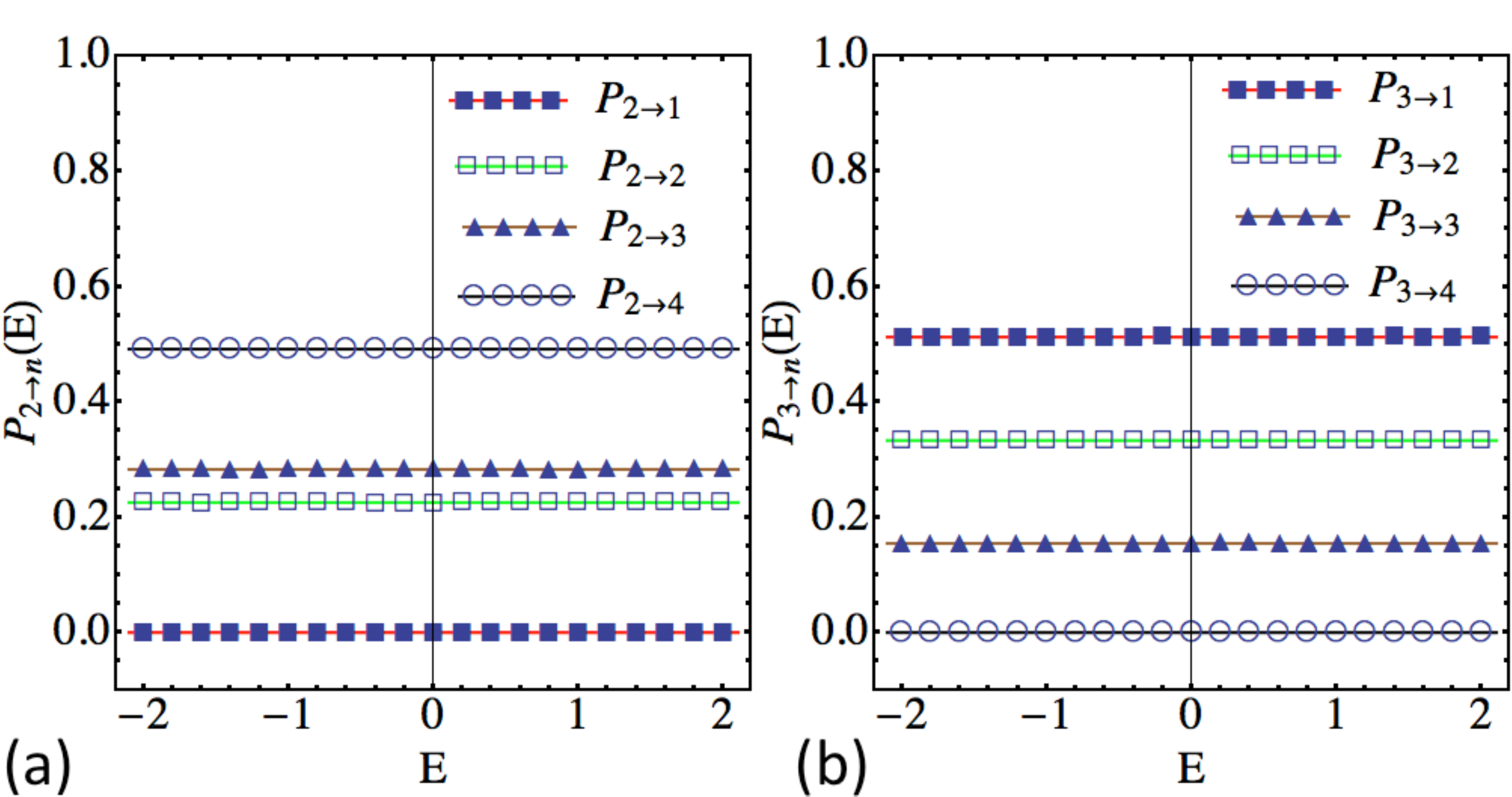}}
\hspace{-2mm}\vspace{-4mm}   
\caption{ (Color online) Numerical test of transition probability independence of the energy bias $E$ for initially populated (a) level-2 and (b) level-3.
All discrete points correspond to results of direct numerical simulations of the evolution with the Hamiltonian (\ref{ham1}) from $t=-600$ to $t=600$. Solid lines are theoretical predictions of Eq.~(\ref{prob1}). The choice of parameters is:  (a) $g=0.45$, $\gamma=0.30$, $\beta_1=0.85$, $\beta_2=0.55$; (b) $g=0.55$, $\gamma=0.35$, $\beta_1=0.95$, $\beta_2=0.65$.}
\label{checkE}
\end{figure}
%%%%%%%%%%%%%%%%%%%%%%%%%%%%%%%%%%%%%%%%%%%%%%%%%%%%%%%%%%%%%%%%%%%%%%%%%%%%%%%%%%%%%%%%%%
So far, our derivation of the transition probability matrix was purely semiclassical. It is expected to be predictive only for sufficiently large values of parameter $E$ that characterizes the separation of the avoided crossing points. 
In order to prove that the matrix (\ref{prob1}) is, in fact, the exact nonperturbative solution of the model with the Hamiltonian (\ref{ham1}) we perform direct numerical simulation of the quantum mechanical evolution of this model. 
The algorithm for such simulations is described in the supplementary file for Ref.~\cite{sinitsyn-13prl}.
First, we test the prediction of Eq.~(\ref{prob1}) that transition probabilities do not depend on parameter $E$, even when it is close to zero so that the independent crossing approximation cannot be justified. 
Figure~\ref{checkE} confirms our expectations. Even when parameter $E$ is substantially smaller than the sizes of coupling constants, transition probabilities agree with the theoretical prediction. We also note that, 
physically, the energy bias $E$ between two quantum dots usually can be tuned by a gate voltage \cite{qd-molecule}, so the prediction of independence of transition probabilities of $E$ can be verified experimentally. 

\begin{figure}%[!htb]
\scalebox{0.22}[0.22]{\includegraphics{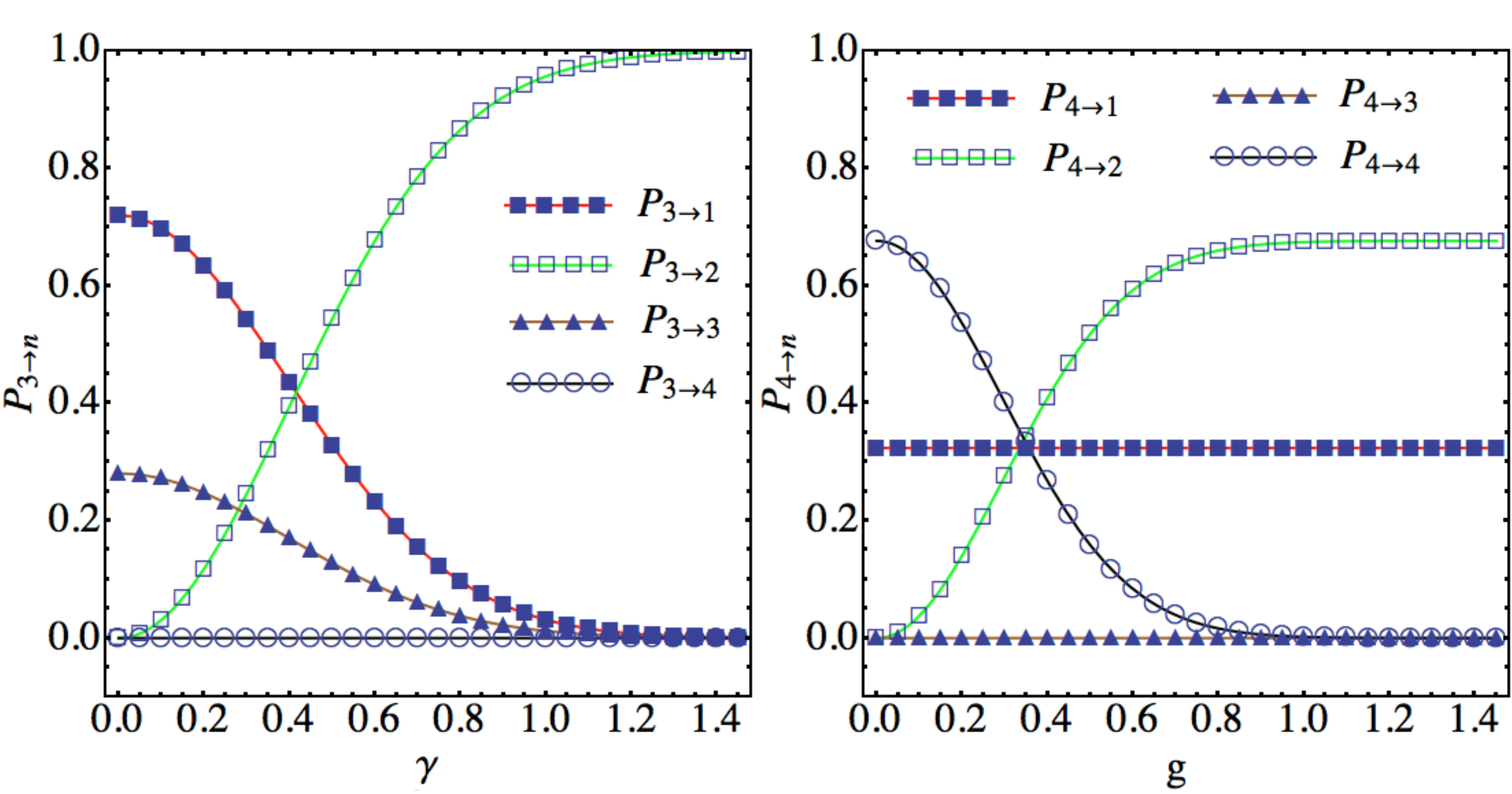}}
\hspace{-2mm}\vspace{-4mm}   
\caption{ (Color online) Numerical check of transition probabilities in Eq.~(\ref{prob1})   (a) as function of $\gamma$ at initially populated level-3; (b) as function of $g$ at  initially populated level-4.
All discrete points correspond to results of direct numerical simulations of the evolution with the Hamiltonian (\ref{ham1}) from $t=-600$ to $t=600$. Solid lines are theoretical predictions of Eq.~(\ref{prob1}). The choice of constant parameters is:  (a) $g=0.45$, $E=-0.5$, $\beta_1=1.0$, $\beta_2=0.5$; (b) $\gamma=0.37$, $E=0.45$, $\beta_1=1.1$, $\beta_2=0.54$.}
\label{check_g}
\end{figure}

In Fig.~\ref{check_g}, we show results of additional tests of Eq.~(\ref{prob1}) at different values of the coupling constants and different initial conditions. In all such tests we found that Eq.~(\ref{prob1}) is in perfect agreement with numerical results, leaving no doubts in that the matrix (\ref{prob1}) is, indeed, the exact nonperturbative solution of the multistate Landau-Zener problem (\ref{ham1}).
  %%%%%%%%%%%%%%%%%%%%%%%%%%%%%%%%%%%%%%%%%%%%%%%%%%%%%%%%%%%%%%%%%%%%%%%%%%%%%%%%%%%%%%%%%%%
%%%%%%%%%%%%%%%%%%%%%%%%%%%%%%%%%%%%%%%%%%%%%%%%%%%%%%%%%%%%%%%%%%%%%%%%%%%%%%%%%%%%%%%%%%

\section{Partial Proof}
Here we show that some of the matrix elements in  Eq.~(\ref{prob1}) and equalities between some of them  can be derived from the discrete symmetry of the model Hamiltonian (\ref{ham1}). Let us introduce  time-dependent amplitudes  of the four diabatic states: $\psi \equiv (a_1(t),a_2(t),a_3(t),a_4(t))^T$. The Schr\"odinger equation (\ref{mlz}) with the Hamiltonian (\ref{ham1}) remains invariant after simultaneous application of three mutually commuting operations:

(a) time reversal $\hat{T}$, i.e. the change of $ t \rightarrow -t$, as well as $a_i \rightarrow a_i^*$, $i=1,2,3,4$;

(b) complex conjugation $\hat{C}$, i.e., the change of the sign near the imaginary unit $i \rightarrow -i$ and replacing  $a_i \rightarrow a_i^*$ or vice versa;

(c) parity operation $\hat{P}$, i.e. renaming the amplitudes $a_1 \rightarrow -a_2$, $a_2 \rightarrow a_1$, and $a_3 \rightarrow -a_4$, $a_4 \rightarrow a_3$. 

The latter operation can be considered as part of the time-reversal operation if we recall that, e.g.,  amplitudes $a_1$ and $a_2$ describe the states of the spin-1/2. However, our definition makes interesting connections with other previously studied multistate Landau-Zener systems, in which such symmetries have been found useful without an obvious spin interpretation of the Hamiltonian \cite{lzc-14pra}. 

After performing those operations,  resulting amplitudes $a_i$ satisfy the same set of ordinary differential equations as the original amplitudes $a_i$. Note that each operation preserves the Hermitian property of the Hamiltonian. Hence, each operation can be reformulated in terms of a transformation of the unitary evolution matrix of the Schr\"odinger equation. Since application of operations (a-c) preserves the form of the evolution equation, the evolution matrix should also remain invariant.

Let $\hat{S}(+\infty| -\infty)$ be the scattering matrix of the model with the Hamiltonian (\ref{ham1}) for evolution from $t=-\infty$ to $t=+\infty$. Time reversal operation converts it to 
$\hat{T} * \hat{S}(+\infty| -\infty) =  \hat{S}(-\infty| +\infty)=\hat{S}^{\dagger} (+\infty| -\infty)$. Complex conjugation corresponds to replacement of elements of the evolution matrix by their complex conjugated values, i.e., $(\hat{C} *\hat{S})_{ij }=S^*_{ij}$. Finally, the parity operation changes places of the matrix elements and adds a proper sign to them, e.g., $(\hat{P}*\hat{S})_{11} = S_{22}$, $(\hat{P} *\hat{S})_{12} = -S_{21}$,   $(\hat{P}* \hat{S})_{13} = S_{24}$,  e.t.c.. Let
\beq
\hat{S} = \left( 
\begin{array}{cccc}
S_{11}&S_{12} &S_{13} &S_{14} \\
S_{21} & S_{22}& S_{23}& S_{24} \\
S_{31}&S_{32} & S_{33} &S_{34}\\
S_{41} & S_{42} &S_{43} & S_{44}
\end{array}
\right)
\label{scatt1}
\eeq
be the original scattering matrix, then
\beq
\hat{S}'\equiv \hat{C} \hat{P} \hat{T} *\hat{S} = \left( 
\begin{array}{cccc}
S_{22}&-S_{12} &S_{42} &-S_{32} \\
-S_{21} & S_{11}& -S_{41}& S_{31} \\
S_{24}&-S_{14} & S_{44} &-S_{34}\\
-S_{23} & S_{13} &-S_{43} & S_{33}
\end{array}
\right).
\label{scatt2}
\eeq
Since $\hat{S} = \hat{S}'$, we can equate corresponding elements of these matrices. For example, comparing elements $S_{12}$ and $S_{12}'$ we find $S_{12}=-S_{12}$, which can be satisfied only if $S_{12}=0$.
Similarly, we find that $S_{12}=S_{21}=S_{34}=S_{43}=0$, from which zero values of the corresponding transition probabilities in Eq.~(\ref{prob1}) follow. 
Comparing elements along the main diagonals of the scattering matrices we find the relations
$S_{22} = S_{11}$, $S_{33}=S_{44}$, e.t.c..  
Let us now write the scattering matrix that includes only elements that cannot be equated to others by a pair-vise comparison of Eqs.~(\ref{scatt1})-(\ref{scatt2}):
\beq
\hat{S} = \left( 
\begin{array}{cccc}
S_{11}&0&S_{13} &S_{14} \\
0 & S_{11}& S_{23}& S_{24} \\
S_{24}&-S_{14} & S_{44} &0\\
-S_{23} & S_{13} &0 & S_{44}
\end{array}
\right).
\label{scatt3}
\eeq
As any unitary matrix, it satisfies the relation 
\beq
\hat{S} \hat{S}^{\dagger}=\hat{1},
\label{unitary1}
\eeq
 where $\hat{1}$ is the unit matrix. Comparing diagonal components of matrices on both sides of this equation we recover the conservation of probability laws, e.g.,
\beq
P_{11}+P_{13}+P_{14}=1, \quad P_{11}+P_{23}+P_{24}=1,
\label{pr1}
\eeq
where we denote $P_{ij}=|S_{ij}|^2$.  Comparing off-diagonal elements of matrices in (\ref{unitary1}), we find relations of the type
\beq
(\hat{S} \hat{S}^{\dagger})_{12}=S_{13}S^*_{23} +S_{14}S^*_{24}=0,
\label{pr2}
\eeq
\beq
(\hat{S} \hat{S}^{\dagger})_{13}=S_{11}S^*_{24} +S_{13}S^*_{44}=0.
\label{pr3}
\eeq
Moving one of the terms in such equations to the right hand side and then  equating the absolute value squared of expressions on both sides we find additional relations between transition probabilities:
\beq
P_{13}P_{23}=P_{14}P_{24}, \quad P_{11}P_{24} =P_{13}P_{44}.
\label{pr4}
\eeq
Substituting (\ref{pr1}) into the first of equations in (\ref{pr4}) we find that $P_{13}=P_{24}$ and substituting this into the 2nd equation in (\ref{pr4}) we find that $P_{11}=P_{44}$. We can now summarize all such results in the following formulas:
\begin{eqnarray}
\label{p-diag}
P_{11}&=&P_{22}=P_{33}=P_{44},\\
P_{13}&=&P_{24}=P_{31}=P_{42},\\
P_{14}&=&P_{23}=P_{32}=P_{41},\\
P_{12}&=&P_{21}=P_{34}=P_{43}=0,
\end{eqnarray}
which are in perfect agreement with (\ref{prob1}). 
This set of equations is also supplemented by the conservation of probabilities, Eq.~(\ref{pr1}), which leaves only two independent transition probabilities undetermined.
%%%%%%%%%%%%%%%%%%%%%%%%%%%%%%%%%%%%%%%%%%%%%%%%%%%%%%%%%%%%%%%%%%%%%%%%%%%%%%%%%%%%%%%%%%%
\begin{figure}%[!htb]
\scalebox{0.24}[0.24]{\includegraphics{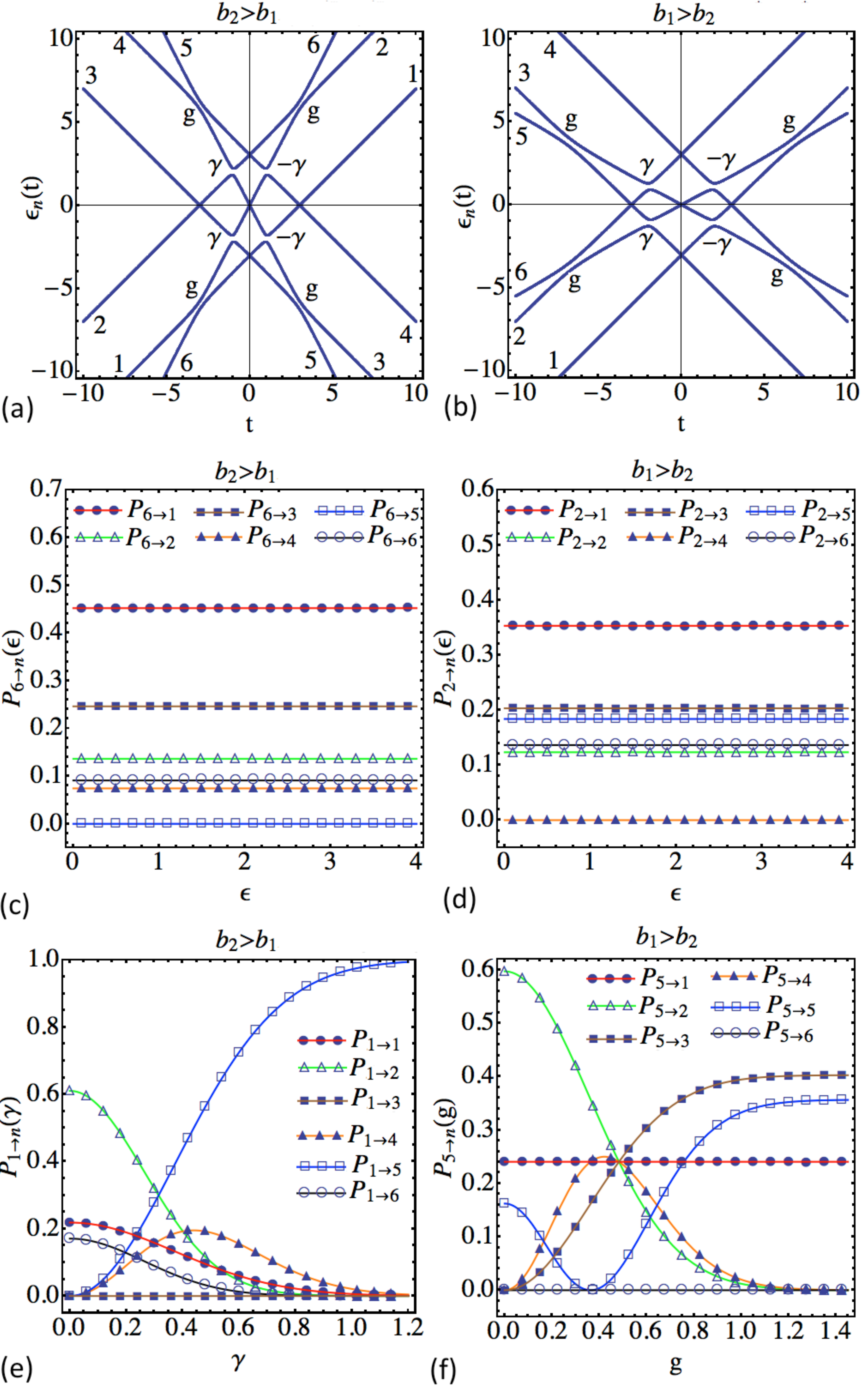}}
\hspace{-2mm}\vspace{-4mm}   
\caption{ (Color online) (a-b) Adiabatic energy levels (blue curves) and their pairwise couplings (parameters $g$ and $\pm \gamma$) of the Hamiltonian (\ref{ham6}) shown at corresponding avoided level crossings. Numbers mark  diabatic levels  that correspond to adiabatic energies at $t\rar  \pm \infty$.  Parameters are:  $\epsilon=3$, $g=0.25$, $\gamma=0.2$ and (a)  $\beta_1=1$, $b_2=0.556$, (b) $b_1=1$, $b_2=1.8$.
(c-d) Tests of independence of transition probabilities of the parameter $\epsilon$ at (c) $b_1=0.75$, $b_2=1.25$, $\gamma=0.3$, $g=0.27$ and (d) $b_1=1.2$, $b_2=0.7$, $\gamma=0.29$, $g=0.38$. (e-f) Tests of 
Eqs.~(\ref{prob2})-(\ref{prob3}) at (e) $\epsilon=0.25$, $b_1=0.25$, $b_2=1.5$, $g=0.55$ and (f) $\epsilon=0.3$, $b_1=1.85$, $b_2=0.24$, $\gamma=0.55$. In (c-f), discrete points correspond to results of numerical simulations of quantum evolution from $t=-800$ to $t=800$ with a time step $dt=0.00001$. Solid curves are  predictions of Eqs.~(\ref{prob2})-(\ref{prob3}).}
\label{levels2}
\end{figure}
%%%%%%%%%%%%%%%%%%%%%%%%%%%%%%%%%%%%%%%%%%%%%%%%%%%%%%%%%%%%%%%%%%%%%%%%%%%%%%%%%%%%%%%%%%%

Until now, our discussion could be equally applied to a more general situation that describes the evolution over any symmetric, around $t=0$, time interval. Moreover, it could be applied not only to linear time-dependence of diagonal matrix elements, e.g, CPT-symmetry is still present if we replace time $t$ in the Hamiltonian (\ref{ham1}) by $t^{\alpha}$, where $\alpha$ is an arbitrary odd integer number. However,  for the multistate Landau-Zener systems of the form (\ref{mlz}), we can achieve more by using some previously derived facts that are specific only for the linear time-evolution of diabatic energies.

If the diabatic level has the extremal slope, then the semiclassical expression for the transition probability to remain at the same level after evolution in $t\in(+\infty, -\infty)$ is known to coincide with its exact value at arbitrary choice of the model parameters. In our case, the extremal slope levels have indexes 1 and 2, so 
\beq
P_{11}=p_1p_2
\label{bef}
\eeq
is the exact result, which is  the consequence of the application of the  Brundobler-Elser formula \cite{be} to the model with the Hamiltonian (\ref{ham1}). This formula was rigorously proved previously \cite{mlz-1} for any model of the type (\ref{mlz}). Hence, combining (\ref{bef}) with (\ref{p-diag}), we obtain all elements along the main diagonal in the matrix (\ref{prob1}) exactly, which also appear in agreement with  the ``semiclassical" prediction.

This leaves only one independent parameter in the transition probability matrix, e.g., the element $P_{14}$, undetermined, while all other facts about Eq.~(\ref{prob1}) follow from the Brundobler-Elser formula and the elementary discrete symmetry of the model. As the CPT-symmetry is insufficient to fix the value of this parameter, we leave the full form of the transition probability matrix (\ref{prob1}) as a conjecture. However, we note  that, in addition to Figs.~\ref{checkE},~\ref{check_g}, 
we performed a series of numerical tests for parameters of the Hamiltonian (\ref{ham1}) beyond the semiclassical and perturbative limits. In all such tests, the deviation of the numerically obtained value of $P_{14}$ 
from its analytical prediction, $P_{14}=q_2$, was found to be below the third significant digit, and it was also within the same accuracy range as the numerically obtained diagonal probabilities $P_{jj}$.  It is likely that the complete
 proof of (\ref{prob1}) can be obtained by the  method that was used previously to prove the Brundobler-Elser  \cite{mlz-1} and no-go \cite{no-go} formulas. 

\section{Extension of the Model}
We showed that the CPT symmetry is at least partially responsible for the integrability of the model (\ref{ham1}). However, the conjecture of integrability (i-ii) appeared more useful here both to recognize the model (\ref{ham1}) as solvable and to obtain its transition probability matrix. 
%In the year 1967, authors of the pioneering article with the solution of the Demkov-Osherov model \cite{do} already pointed that the exact solution of their model reproduced the prediction of the independent crossing approximation. This fact was repeatedly observed in all publications that reported new exact results in the multistate Landau-Zener theory. Moreover, in several cases exact results were anticipated based on the semiclassical intuition and only later such conjectures were confirmed analytically \cite{bow-tie,be,no-go}. In our recent publication \cite{six-LZ}, we suggested that the conditions (i-ii) are likely the signature for existence of a class of exactly solvable models with properties (i-ii). 

While currently there is no known algorithm to generate models with these properties, we did a trial and error search for conditions (i-ii) among systems that have  similar level crossing patterns to the model (\ref{ham1}). One example that we found describes the system of 6-states:
\begin{widetext}
\begin{equation}
\hat{H}=\left( 
\begin{array}{cccccc}
b_1 t -\epsilon & 0 & 0 & 0 & -\gamma & g \\
0 & b_1 t +\epsilon & 0 & 0 & \gamma & g  \\
0 & 0 & -b_1 t-\epsilon & 0 &g & \gamma \\
0& 0& 0& -b_1 t +\epsilon & g & -\gamma \\
-\gamma & \gamma & g & g & -b_2 t & 0\\
g&g & \gamma & -\gamma & 0 & b_2 t
\end{array}
\right).
\label{ham6}
\end{equation}
\end{widetext}
This model  can also be interpreted in terms of an electron hopping between two quantum dots. One of the dots, in this case, has a more complex structure. At zero magnetic field, instead of a single spin-degenerate quantum level with energy $E$, it has  two spin-degenerate quantum levels with energy distances  $\pm \epsilon$ from the level of the other dot.   
Figures~\ref{levels2}(a-b) show the adiabatic energies of the Hamiltonian (\ref{ham6}) as functions of $t$. This time, there are five exact crossing points: three at $t=0$ and additional two at zero values of diabatic energies, which are the consequence of additional discrete symmetries of the system. 
Let us denote
\beq
p_1\equiv e^{-\frac{2\pi g^2}{|b_1-b_2|}}, \quad p_2 \equiv  e^{-\frac{2\pi \gamma^2}{b_1+b_2}}, \quad q_n\equiv 1-p_n.
\label{pq-2}
\eeq
The independent crossing approximation makes different predictions for transition probabilities, depending on whether $b_1>b_2$ or $b_1<b_2$:
\begin{widetext}
\begin{equation}
\hat{P}^{(b_2>b_1)}=\left( 
\begin{array}{cccccc}
 p_1p_2& q_2^2 & 0 &  p_2q_1q_2& p_1p_2q_2 &  p_2q_1\\
(p_2q_1)^2 &  p_1p_2& p_2q_2q_1 &0  & q_2 &  p_2^2p_1q_1 \\
 0& p_2q_2q_1 & p_1p_2&  (p_2q_1)^2 &p_2^2p_1q_1 &  q_2\\
p_2q_2q_1& 0& q_2^2& p_1p_2 & q_1p_2 & p_1p_2q_2 \\
 q_2&  p_1p_2q_2& p_2q_1 & p_2^2p_1q_1 & (p_1p_2)^2 &0\\
p_2^2p_1q_1&p_2q_1 &  q_2p_2p_1& q_2 & 0 & (p_1p_2)^2
\end{array}
\right),
\label{prob2}
\end{equation}
\begin{equation}
\hat{P}^{(b_1>b_2)}=\left( 
\begin{array}{cccccc}
 p_1p_2&  (p_1-p_2)^2& 0 & q_1q_2 & p_2q_2&  p_1q_1\\
 0&  p_1p_2& q_1q_2 & 0 &p_1q_2&  p_2q_1 \\
 0& q_1q_2 & p_1p_2& 0  &p_2q_1 &  p_1q_2\\
q_1q_2& 0& (p_1-p_2)^2& p_1p_2 & p_1q_1&p_2q_2\\
 p_1q_2& p_2q_2 & p_1q_1 &  p_2q_1& (p_1+p_2-1)^2 &0\\
p_2q_1& p_1q_1 &p_2q_2  & p_1q_2& 0 & (p_1+p_2-1)^2
\end{array}
\right).
\label{prob3}
\end{equation}
\end{widetext}
In Figs.~\ref{levels2}(c-f) we test predictions of Eqs.~(\ref{prob2})-(\ref{prob3}) in the regime beyond the semiclassical limit and find perfect agreement of numerical and analytical results. This observation leaves practically no doubts that the models (\ref{ham1}) and (\ref{ham6}) are exactly solvable, and  conditions (i-ii) can be used as a guide to search for exact results in the multistate Landau-Zener theory. Moreover, considering the existence of a solvable 6-state system with similar symmetries \cite{six-LZ}, it is likely that the models (\ref{ham1}), (\ref{ham6}) and the 6-state model, discussed in \cite{six-LZ}, are the instances of a bigger class of  solvable systems. There are hints that may become useful in search for this class. All its found instances have multiple exact crossing points of adiabatic levels at $t=0$, which suggests that models of this class describe fermionic systems under the action of a time-dependent magnetic field, with time-reversal invariance at zero magnetic field value. We also note that  the presented models are similar in structure to the reducible 4- and 6-state models described in \cite{multiparticle}, which points to a possible duality between these systems or existence of a similar algebraic structure responsible for integrability.  We would also like to mention a different view on integrability of multistate Landau-Zener systems that was proposed recently in \cite{armen}. It relates presence of exact adiabatic level crossings with existence of nontrivial commuting Hamiltonians at arbitrary time values \cite{com-partner}.

\section{Conclusion}

We identified and explored  4- and 6-state Landau-Zener models, for which we determined the exact form of the matrix of transition probabilities. These models show a relatively complex behavior due to the possibility of semiclassical path interference leading to either constructive or destructive interference.  Our numerical simulations and the partial proof confirm that the transition probability matrices, which we derived in a semiclassical framework, are exact, i.e., they describe arbitrary choices of the model parameters.  

Not all exact results in the multistate Landau-Zener theory have found practical applications. For example, the generalized bow-tie model \cite{bow-tie},  having been mathematically influential, looks quite artificial from the practical point of view, and its physical realization is unknown. Our result shows that there are solvable multistate Landau-Zener systems that can be of a practical interest to the on-going experimental studies. One useful property of the solution (\ref{prob1}) is that some of the transition probabilities depend only on the parameter $\gamma$ that characterizes the spin-orbit coupling. This suggests a simple way to extract this coupling constant experimentally. Another property is the existence of completely destructive interference for some transitions, which should be strongly sensitive to decoherence.

Finally, we argued that the considered models are likely only special instances of a more general class of integrable systems. It would be highly desirable to find a general Hamiltonian for  this class in order to obtain an exact framework to explore complex interactions in nonstationary quantum mechanics. 

{\it Acknowledgment}. Author thanks Rolando Somma for useful discussions. The work
was carried out under the auspices of the National Nuclear
Security Administration of the U.S. Department of Energy at Los
Alamos National Laboratory under Contract No. DE-AC52-06NA25396. Author also thanks the support from the LDRD program at LANL.


\begin{thebibliography}{100}

\bibitem{dot-lz-exp1} G. Cao, {\it et al.}, Nature Comm. {\bf 4}, 1401 (2012)
\bibitem{multiparticle} N. A. Sinitsyn, Phys. Rev. B {\bf 66}, 205303 (2002)


\bibitem{dot-lz-exp2} J. M. Nichol, {\it et al.}, Nature Comm. {\bf 6} 7682 (2015);  C. Zhou, Phys. Rev. A {\bf 89}, 022337 (2014); L. Gaudreau, Nature Phys. {\bf 8}, 54 (2012)
\bibitem{dot-lz-theory1} P. Nalbach, J. Kn\"orzer and S. Ludwig, Phys. Rev. B {\bf 87}, 165425 (2013); H. Ribeiro, J. R. Petta, and G. Burkardn, Phys. Rev. B {\bf 87}, 235318 (2013); H. Ribeiro, G. Burkard,  Phys. Rev. Lett. {\bf 102}, 216802 (2009)

%\bibitem{app-spin} W. Wernsdorfer {\it et al}, J. Appl. Phys. {\bf 87}, 5481 (2000), W. Wernsdorfer {\it et al}, Phys. Rev. Lett. {\bf 84}, 2965 (2000); W. Wernsdorfer {\it et al}, Europhys. Lett., {\bf 50} (4),  552 (2000); N. A. Sinitsyn and N. Prokof'ev, Phys. Rev. B {\bf 67}, 134403 (2003)

%\bibitem{atomic} Y-A. Chen, S. D. Huber,  S. Trotzky,	 I. Bloch, and E. Altman, Nature Phys. 7, 61 (2011); C. Kasztelan, S. Trotzky, Y.-A. Chen, I. Bloch, I. P. McCulloch, U. Schollw\"ock, and G. Orso, Phys. Rev. Lett. 106, 155302 (2011); S F. Caballero-Benitez and R. Paredes, Phys. Rev. A 85, 023605 (2012); W. Tschischik, M. Haque, R. Moessner, Phys. Rev. A 86, 063633 (2012)

%\bibitem{LZ-interferometry} M. N. Kiselev, K. Kikoin and M. B. Kenmoe, EPL {\bf 104}, 57004 (2013); F. Forster {\it et al} Phys. Rev. Lett. {\bf 112}, 116803 (2014);  Sriram Ganeshan, Edwin Barnes, and S. Das Sarma, Phys. Rev. Lett. {\bf 111}, 130405 (2013); Hugo Ribeiro, J. R. Petta, and G. Burkard, Phys. Rev. B {\bf 87}, 235318 (2013)


%\bibitem{qcontrol} C. M. Quintana, K. D. Petersson, L. W. McFaul, S. J. Srinivasan, A. A. Houck, J. R. Petta, Phys. Rev. Lett. {\bf 110}, 173603 (2013); S. Masuda, K. Nakamura, and A. del Campo, Phys. Rev. Lett. {\bf 113}, 063003 (2014); S. Deffner, C. Jarzynski, and A. del Campo, Phys. Rev. X {\bf 4}, 021013 (2014); A. del Campo, M. M. Rams, and W. H. Zurek, Phys. Rev. Lett. {\bf 109}, 115703 (2012); J. Keeling, A. V. Shytov, and L. S. Levitov, Phys. Rev. Lett. {\bf 101}, 196404 (2008)

%\bibitem{coher} K. Saito, M. Wubs, S. Kohler, Y. Kayanuma, and P. H\"anggi,
%Phys. Rev. B {\bf 75}, 214308 (2007); P. Ao and J. Rammer, Phys. Rev. B {\bf 43}, 5397 (1991) ; M. H. S. Amin, D. V. Averin, and J. A. Nesteroff, Phys. Rev. A {\bf 79}, 022107 (2009); V. N. Ostrovsky and M. V. Volkov,
%Phys. Rev. B {\bf 73}, 060405 (2006);  M. Wubs, K. Saito, S. Kohler, P. H\"anggi, and Y. Kayanuma, Phys. Rev. Lett. {\bf 97}, 200404 (2006).

%\bibitem{app-bose}  V. A. Yurovsky, A. Ben-Reuven, and P. S. Julienne, Phys. Rev. A {\bf 65}, 043607 (2002);   V Shahnazaryan, O Kyriienko, I Shelykh, Preprint arXiv/1410.1379 (2014);
%W. H. Zurek, U. Dorner, and P. Zoller,
%Phys. Rev. Lett. {\bf 95}, 105701 (2005);  B. Damski and W. H. Zurek
%Phys. Rev. A {\bf 73}, 063405 (2006);  B. Damski, H. T. Quan, and W. H. Zurek, Phys. Rev. A {\bf 83}, 062104 (2011); V. Gurarie, Phys. Rev. A {\bf 80}, 023626 (2009);  B. E. Dobrescu, and V. L. Pokrovsky, Phys. Letters A {\bf 350},  154 (2006); M. Schecter, and A. Kamenev, Phys. Rev. A {\bf 85}, 043623 (2012)


%\bibitem{maj} E. Majorana, Nuovo Cimento {\bf 9} (2), 43 (1932); L. D. Landau, Physik Z. Sowjetunion {\bf 2}, 46 (1932);  C. Zener, Proc. R. Soc. A {\bf 137}, 696 (1932); 

\bibitem{book} H. Nakamura, ``Nonadiabatic Transition", World Scientific Publishing Company, 2nd Edition (2012); M. S. Child, ``Molecular Collision Theory", Dover Publications (2010); E. E. Nikitin, S. Y. Umanskii, ``Theory of Slow Atomic Collisions", Springer Series in Chemical Physics (Book 30), Springer (2011)


\bibitem{be}S. Brundobler and V. Elser, J. Phys. A {\bf 26}, 1211 (1993)


%\bibitem{landau}  L. D. Landau, Physik Z. Sowjetunion {\bf 2}, 46 (1932)
%\bibitem{LZ} C. Zener, Proc. R. Soc. A {\bf 137}, 696 (1932)
%\bibitem{stuck} E. C. G. St\"uckelberg, Helv. Phys. Acta {\bf 5}, 369 (1932)

%\bibitem{sinitsyn-14pra} N. A. Sinitsyn, Phys. Rev. A {\bf 90}, 062509 (2014)



%\bibitem{shytov}  A. V. Shytov, Phys. Rev. A {\bf 70}, 052708 (2004)



%A. V. Shytov, Phys. Rev. A {\bf 70}, 052708 (2004); 

\bibitem{do} Yu. N. Demkov and V. I. Osherov, Zh. Exp. Teor. Fiz. {\bf 53}, 1589 (1967) [Sov. Phys. JETP {\bf 26}, 916 (1968)]; A. A. Rangelov, J. Piilo, and N. V. Vitanov, Phys. Rev. A {\bf 72}, 053404 (2005)


%\bibitem{reducible} J. Dziarmaga, Phys. Rev. Lett. {\bf 95}, 245701 (2005);  M. V. Volkov and V. N. Ostrovsky, Phys. Rev. A {\bf 75}, 022105 (2007)

\bibitem{bow-tie}  Y. N. Demkov and V. N. Ostrovsky, Phys. Rev. A {\bf 61}, 032705 (2000)
%\bibitem{bow-tie1} Yu. N. Demkov and V. N. Ostrovsky, J. Phys. B {\bf 28}, 403 (1995); V. N. Ostrovsky and H. Nakamura, J. Phys. A {\bf 30}, 6939 (1997);  Y. N. Demkov and V. N. Ostrovsky, J. Phys. B {\bf 34}, 2419 (2001); C. E. Carroll and F. T. Hioe, J. Phys. A {\bf 19}, 1151 (1986)

%\bibitem{chain}  N. A. Sinitsyn, Phys. Rev. A {\bf 87}, 032701 (2013); V. L. Pokrovsky and N. A. Sinitsyn, Phys. Rev. B {\bf 65}, 153105 (2002)




\bibitem{six-LZ} N. A.  Sinitsyn, J. Phys. A: Math. Theor. {\bf 48} 195305 (2015) 


\bibitem{armen} A. Patra, E. A. Yuzbashyan, J. Phys. A: Math. Theor. {\bf 48}, 245303 (2015)

\bibitem{roy} D. Roy, Yan Li, A. Greilich, Yu. V. Pershin, A. Saxena, and N. A. Sinitsyn, Phys. Rev. B {\bf 88}, 045320 (2013)

\bibitem{qd-molecule} A. Greilich \emph{et al.}, Nature Photonics {\bf 5}, 702 (2011)

\bibitem{inverseFA-fast} A. V. Kimel, et al, Nature Lett. {\bf 435}, 655 (2005)

\bibitem{inverseFA-dirac} E. J. Sie, J. W. McIver, Y. Lee, L. Fu,	J. Kong and N. Gedik, Nature Materials {\bf 14}, 290Ð294 (2015)

\bibitem{sinitsyn-13prl} N. A. Sinitsyn, Phys. Rev. Lett. {\bf 110}, 150603 (2013) 

\bibitem{lzc-14pra} N. A. Sinitsyn, Phys. Rev. A {\bf 90}, 062509 (2014)

\bibitem{mlz-1}     M. V. Volkov and V. N. Ostrovsky, J. Phys. B: At. Mol. Opt. Phys. {\bf 37}, 4069 (2004);  B. E. Dobrescu and N. A. Sinitsyn, J. Phys. B: At. Mol. Opt. Phys. {\bf 39}, 1253 (2006)



\bibitem{no-go} N. A. Sinitsyn,  J. Phys. A {\bf 37} (44), 10691 (2004); M. V. Volkov and V. N. Ostrovsky,  J. Phys. B: At. Mol. Opt. Phys. {\bf 38}, 907 (2005)

\bibitem{com-partner} E. A. Yuzbashyan, B. S. Shastry, J. Stat. Phys. {\bf 150}, 704 (2013)

\end{thebibliography}
\end{document}